\documentclass[journal,onecolumn]{IEEEtran}
\usepackage{cite}
\ifCLASSINFOpdf
\else
\fi

\usepackage[cmex10]{amsmath}
\usepackage{array}
\usepackage{mdwmath}
\usepackage{mdwtab}
\usepackage{xcolor}

\usepackage{graphicx}

\usepackage{amsmath,amsfonts,amssymb}% better equation formatting
\usepackage{tabularx}

\usepackage{epstopdf} % When using eps figures 

\usepackage{hyperref}

\usepackage{caption}
\usepackage{subcaption}

\DeclareMathOperator{\acos}{acos}
\DeclareMathOperator{\asin}{asin}

\DeclareMathOperator*{\argmax}{arg\,max}

\newcommand\tinyEarth{\vcenter{\hbox{\scalebox{0.5}{$\oplus$}}}}
\newcommand\LoS{\vcenter{\hbox{\scalebox{0.5}{$\mathrm{LoS}$}}}}
\newcommand\NLoS{\vcenter{\hbox{\scalebox{0.5}{$\mathrm{NLoS}$}}}}

\begin{document}

\title{Optimal Beamwidth and Altitude for\\ Maximal Uplink Coverage in Satellite Networks}

\author{Bassel Al Homssi,~\IEEEmembership{Member,~IEEE}, and Akram~Al-Hourani,~\IEEEmembership{Senior Member,~IEEE}.
	
	\thanks{B. Al Homssi and A. Al-Hourani are with the School of Engineering, RMIT University, Melbourne, Australia. E-mail: bhomssi@ieee.org and akram.hourani@rmit.edu.au.}
}

\maketitle

\begin{abstract}
	Dense satellite constellations recently emerged as a prominent solution to complementing terrestrial networks in attaining true global coverage. As such, analytic optimization techniques can be adopted to rapidly maximize the benefits of such satellite networks. The paper presents a framework that relies on two primary tuning parameters to optimize the uplink performance; (i) the constellation altitude and (ii) the satellite antenna beamwidth. The framework leverages tools from stochastic geometry to derive analytical models that formulate a parametric uplink coverage problem which also includes user traffic demand as an input. This allows operators to devise uplink expansion strategies to cater for expanding user demand. The framework demonstrates that fine-tuning of these parameters can significantly enhance the network capacity. We show that the optimization of random constellations provides a close match to that of practical satellite constellations such as Walker-delta and Walker-star.
\end{abstract}


\IEEEpeerreviewmaketitle 

\section{Introduction}
\IEEEPARstart{D}{riven} by aspirations to attain continuum hybrid satellite-terrestrial coverage, dense satellite constellations recently emerged as an appealing solution for next generation networks. Accordingly, thousands of satellites are currently being deployed to provide coverage to ground users in areas with limited terrestrial infrastructure~\cite{1522108}. Applications that rely on massive Internet-of-Things (IoT) networks are becoming increasingly more diverse. Many of these applications rely on capturing massive information from physical phenomena, e.g. location, humidity, and temperature, and then upload this information through the IoT access network. However, adequate terrestrial coverage is usually limited for urbanized regions leaving the rural and offshore applications unsupported. Thereby satellite constellations is an efficient and feasible alternative to provide connectivity for these applications~\cite{de2015satellite}.

These constellations are expected to operate in low Earth orbit (LEO) constellations due to their short propagation distance and thus having lower latency~\cite{8002583}. Due to the inherent nature of LEO, the satellite footprint is confined due to Earth's occlusion. While higher satellite availability is typically desirable, a bigger footprint contributes to higher aggregated uplink interference. As such, constellation altitude is one of the parameters that impacts the performance of the communication link~\cite{9390220}. Alternatively, coverage beamwidth can be controlled to artificially limit the interference region. Both the altitude and beamwidth can be chosen to tune the uplink performance where a trade-off occurs between the availability and interference. As a result, a network design that takes into account the optimal parameters is imminent to capitalize on the network's performance.

Moreover, IoT-over-satellite networks enable massive devices to join the network, which aggravates the interference problem even further and leads to degradation in coverage performance~\cite{7476821,9395074}. Thus, expansion strategies are imminent for the network to be robust and consequently maintain the required quality-of-service (QoS). Conventionally, specialized software platforms that provide tailored simulations are deployed in order to obtain the performance of geostationary satellite links. In the case of massive satellite constellations, the complexity of those simulations increases and is especially worsened for network optimization applications where iterations are required. Alternatively, analytic approaches that rely on stochastic geometry, provide deep insights into the different parameters impacting the performance of the complex networks~\cite{6042301,5226957}. Examples of analytic models that capture the performance for massive satellite constellations are derived~\cite{9313025,9218989,9177073}. Some expansion strategies have focused on maximizing the downlink coverage using the constellation altitude~\cite{9390220}. However in instances where the network is already deployed, changing the satellite altitude is an infeasible approach, making altitude tuning insufficient on its own. 

In this paper, we develop an analytic framework that maximizes the uplink coverage probability for dense satellite constellations by jointly tuning both the altitude and beamwidth. The framework capitalizes on tools from stochastic geometry to obtain analytical modeling for the uplink coverage probability that takes into account the ground user traffic demand which is especially useful for networks that rely on massive users such as IoT-over-satellite. This framework facilitates the strategic deployment and expansion of satellite infrastructure to cater for increasing user demand. The optimization results show a close match with practical constellations based on Walker-delta and Walker-star.

\begin{figure*}
	\captionsetup{font=footnotesize,labelfont=footnotesize,justification=raggedright,singlelinecheck = false}
	\begin{subfigure}{0.325\linewidth}
		\caption{Occlusion-limited, $\psi = \psi_\mathrm{o}$}
		\includegraphics[width=\linewidth]{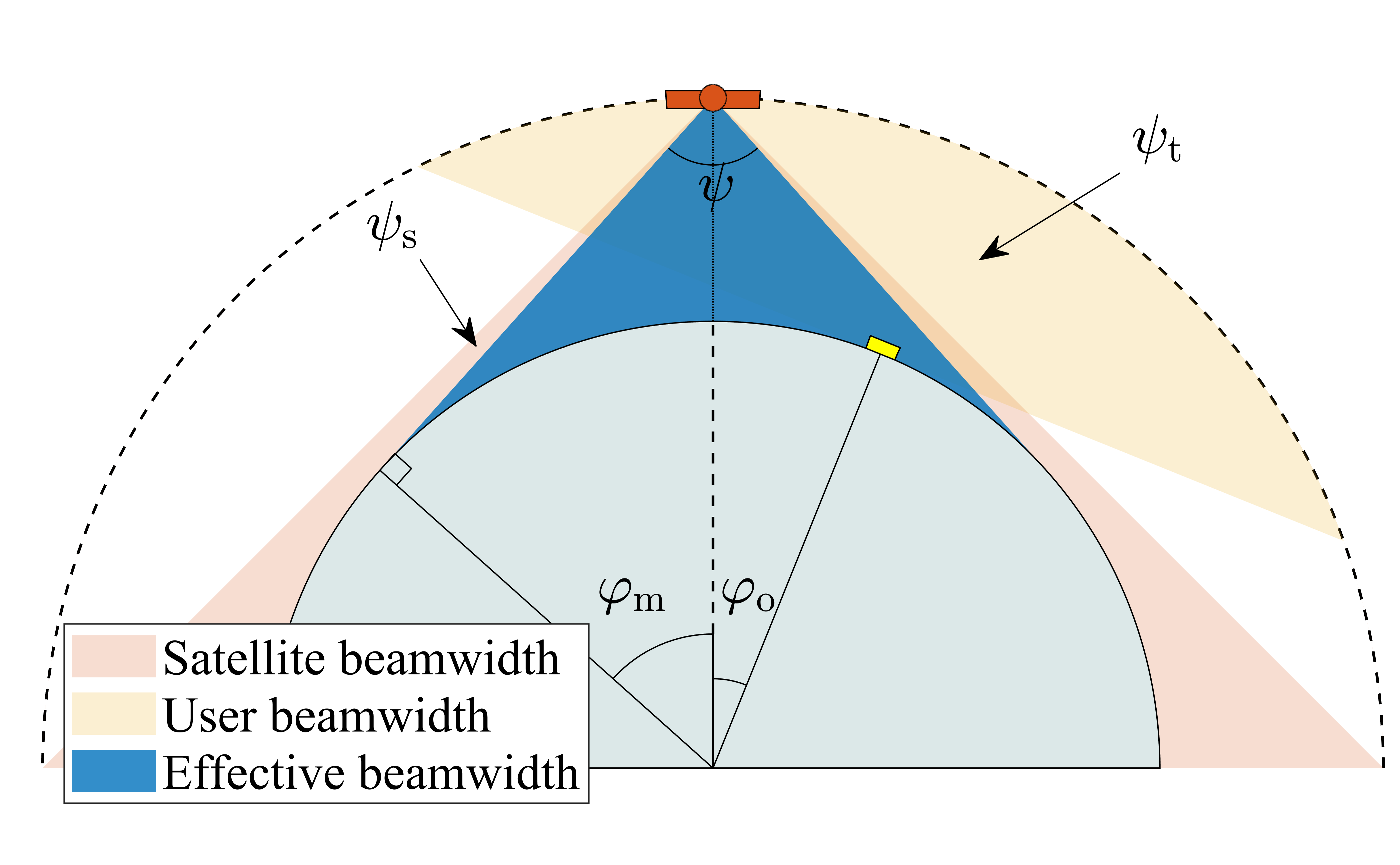}
	\end{subfigure}
	\begin{subfigure}{0.325\linewidth}
		\caption{Satellite-limited, $\psi = \psi_\mathrm{s}$}
		\includegraphics[width=\linewidth]{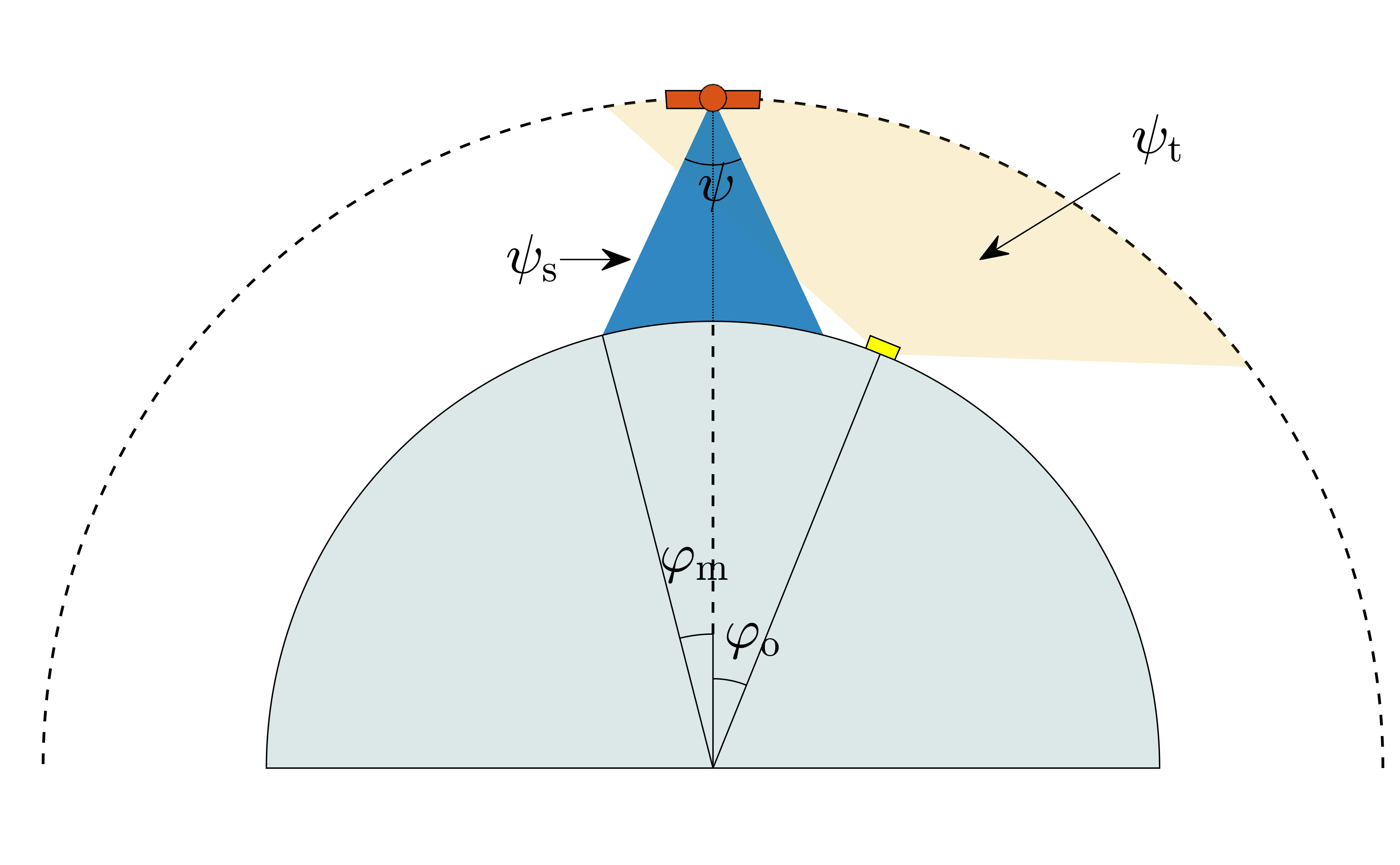}
	\end{subfigure}
	\begin{subfigure}{0.325\linewidth}
		\caption{Device-limited, $\psi = 2\asin\left[\alpha\sin(\psi_\mathrm{t}/2)\right]$}
		\includegraphics[width=\linewidth]{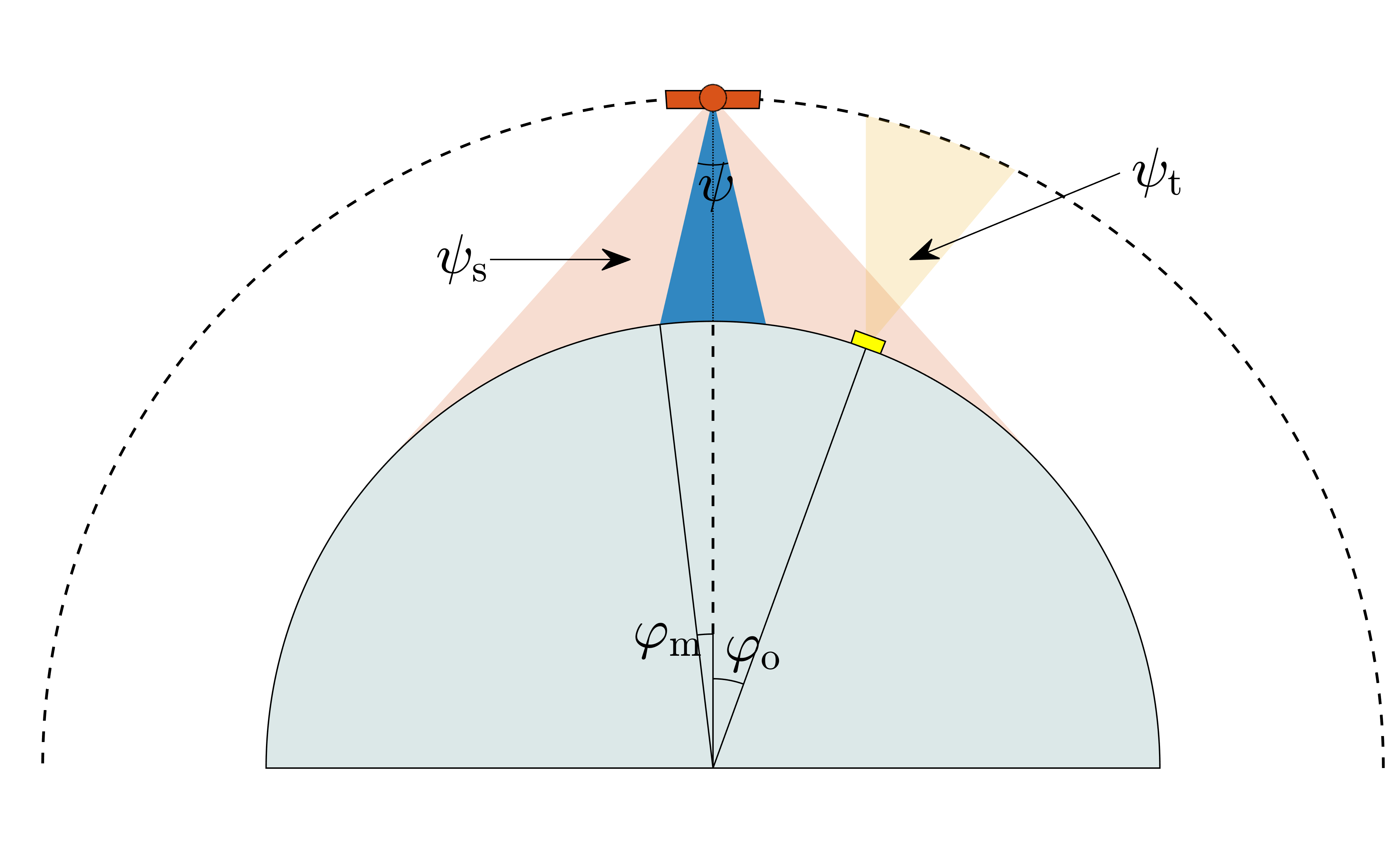}
	\end{subfigure}
	\centering
	\caption{A satellite-centric representation showcasing the effective beamwidth generated from combining both satellite and ground user beamwidths.}
	\label{Fig_Schematic}
\end{figure*}
\section{System Model}\label{Sec_Model}

\subsection{Geometric Model}
\noindent
Consider a massive constellation of $N$ satellites with circular orbits at an altitude $h$ above the mean sea level. In order to facilitate tractability, we assume that these satellites are rotating at random orbital such that the locations of the satellites are uncorrelated at any given moment. From the user's perspective, the satellites can be approximated using the homogeneous Binomial Point Process (BPP). Moreover, the active user locations are also assumed to be homogeneously distributed on the Earth's surface and are represented with another independent BPP with a constant density, $\lambda = D\lambda_\mathrm{o}$, where $\lambda_\mathrm{o}$ is the density of all ground user devices and $D$ is the spatial duty cycle. Note that Earth is assumed to be an ideal sphere with an average radius denoted as $R_{\tinyEarth}$.

In order for a user to communicate with a given satellite, a set of conditions need to be met; (i)	the user location needs to be within the horizon of the satellite, (ii) the ground user needs to be within the region covered by the satellite footprint governed by its beamwidth, and (iii) the satellite needs to be within the antenna beam of the user. Accordingly, there is an \textit{effective beamwidth}, denoted as $\psi$, which corresponds to the minimum satellite-centric angle at which the satellite can provide connectivity. From the satellite perspective, there exists a section of Earth's surface, basically a spherical cap, representing the effective footprint. This spherical cap is bounded by an Earth-centered apex angle of $2\varphi_\mathrm{m}$, whereas the angle $\varphi_\mathrm{m}$ is the defined as the \textit{maximum Earth-centered zenith angle}. The different scenarios that impact this effective footprint is depicted in Fig.~\ref{Fig_Schematic}. Using mathematical reasoning, the effective beamwidth and the maximum zenith angles are given as follows
\begin{equation}
	\psi = \min\left[\psi_\mathrm{s}, 2\asin\left(\alpha\sin\frac{\psi_\mathrm{t}}{2}\right)\right] ,
\end{equation}
where $\psi_\mathrm{s},\psi_\mathrm{t}$ are the beamwidth of the satellite and user, respectively, and $\alpha = R_{\tinyEarth}/(R_{\tinyEarth} + h)$. 
The term $2\asin\left(\alpha\sin\frac{\psi_\mathrm{t}}{2}\right)$ is the satellite-centric beamwidth that is equivalent to a ground user beamwidth, derived using the law of sines. The maximum Earth-centered Zenith is obtained as follows
\begin{equation}
	\varphi_\mathrm{m} = 
	\begin{cases}
		\asin\left(\frac{1}{\alpha}\sin \frac{\psi}{2}\right) - \frac{\psi}{2} & , \psi<\psi_\mathrm{o} \\
		\acos \alpha &, \psi>\psi_\mathrm{o}
	\end{cases},
\end{equation}
where $\psi_\mathrm{o}$ is the satellite beamwidth that is just covering the horizon, i.e., $\psi_\mathrm{o} = 2\asin\alpha$. As a result, the effective footprint is predicated on (i) the altitude of the constellation, (ii) the satellite antenna beamwidth, and (iii) the ground user antenna beamwidth. In practical scenarios, controlling the effective footprint can be achieved by either varying $\psi_\mathrm{s}$ and $\psi_\mathrm{t}$ for a given altitude.

In this model, we assume that each user associates with its nearest satellite, i.e., satellite that is closest to the user's zenith angle~\cite{9313025}. We define the \textit{contact angle} as the Earth-centered zenith angle of this closest satellite, $\varphi_\mathrm{o}$. The satellite availability is the probability that the target user has at least one satellite is capable of providing it with service. The availability can be described using the contact angle cumulative density function (CDF), whereby ${\mathbb{P}(\varphi_\mathrm{o}<\varphi) = F_{\varphi_\mathrm{o}}(\varphi)}$,  obtained as follows~\cite{9313025}
\begin{equation}\label{eq_availabilitycdf}
	F_{\varphi_\mathrm{o}}(\varphi) = 1 - \exp\left(-\frac{N}{2}[1-\cos\varphi]\right)~.
\end{equation}
Moreover, the probability density function (PDF) of the contact angle is obtained as follows
\begin{equation}\label{eq_availabilitypdf}
	f_{\varphi_\mathrm{o}}(\varphi) = \frac{N}{2}\sin\varphi\exp\left(-\frac{N}{2}[1-\cos\varphi]\right)~.
\end{equation}
where $\varphi_\mathrm{o}$ is the contact angle. Note that in dense constellations that have a considerably large $N$, ground users can still be within a satellite footprint and contribute to its interference even if this satellite isn't serving them.

\subsection{Ground-to-Satellite Channel Model}
We consider that all ground users are transmitting with the same transmit power, denoted $P_\mathrm{t}$. Accordingly, the received power at the satellite is given by
\begin{equation}\label{Eq_ReceivedPower}
	P_\mathrm{r} = P_\mathrm{t}G_{\mathrm{s}}G_{\mathrm{t}}l~\zeta,
\end{equation}
where $G_{\mathrm{s}}$ and $G_{\mathrm{t}}$ are the gains of the satellite and user ground antennas, respectively, and $l$ is the free space path-gain expressed in terms of the zenith angle as follows
\begin{equation}
	l = \left(\frac{c}{4\pi f}\right)^2\frac{1}{R_{\tinyEarth}^2 + (R_{\tinyEarth} + h)^2 - 2R_{\tinyEarth}(R_{\tinyEarth}+h)\cos\varphi},
\end{equation}
where $c$ is the speed of light and $f$ is the carrier frequency. $\zeta$ in \eqref{Eq_ReceivedPower} is the excess-gain, expressed in terms of a Gaussian mixture model (GMM) as follows~\cite{9257490}
\begin{equation}
	\zeta[\mathrm{dB}] \sim p_{\LoS}(\varphi)\mathcal{N}(-\mu_{\LoS},\sigma_{\LoS}^2)  + p_{\NLoS}(\varphi)\mathcal{N}(-\mu_{\NLoS},\sigma_{\NLoS}^2) ,
\end{equation}
where $p_{\LoS}$ is the probability of line-of-sight (LoS) and its complement, $p_{\NLoS}$ is the probability of non-line-of-sight (NLoS). The LoS probability is obtained as follows
\begin{equation}
	p_{\LoS}(\varphi) = \exp\left(-\frac{\beta\sin\varphi}{\cos\varphi - \alpha}\right) ,
\end{equation}
where $\beta, \mu_{\LoS}, \sigma_{\LoS}, \mu_{\NLoS}, \sigma_{\NLoS}$ depend on the propagation environment~\cite{9257490}.
\vspace{-5.5mm}
\subsection{Probability of Coverage}
The aggregate interference at the satellite receiver is the summation of all active users located within the satellite footprint, given as follows
\begin{equation}\label{eq_ISummation}
	I = \sum_{x_i\in\Phi\cap\mathcal{A}\backslash x_\mathrm{o}} \kappa P_\mathrm{t}G_\mathrm{t}G_\mathrm{s}l(\varphi_i)\zeta(\varphi_i)~,
\end{equation}
where $\kappa \in [0, 1]$ is the interference mitigation factor that captures the effect of the satellite access system in scheduling shared radio resources, where $\kappa = 0$ indicates an ideal scheduling system with no co-channel interference, whereas $\kappa = 1$ represents a complete random access system. The summation is performed over the footprint region denoted as $\mathcal{A}$. When considering a large footprint, the number of interfering devices converges to the average value of $\lambda||\mathcal{A}||$. Accordingly, the variations in the interference are considered negligible when compared to the magnitude of the average interference. As such, the average can be reliably used to describe the interference level~\cite{9509510}. To simplify our analysis, we assume that both the ground user and satellite antenna gains are uniform and are thus independent of the boresight angle. The average interference is derived as follows
\begin{equation}\label{Eq_Ibar}
	\bar{I} = 2\pi \lambda R_{\tinyEarth}^2 \kappa P_\mathrm{t}G_\mathrm{t}G_\mathrm{s} \int_{0}^{\varphi_\mathrm{m}}  l(\varphi)\bar{\zeta}(\varphi)\sin\varphi~\mathrm{d}\varphi~.
\end{equation}
For the detailed derivation, see Appendix~\ref{Appendix_a}. For a GMM excess path-gain, the average value $\bar{\zeta}$ in linear form is given as follows
\begin{equation}
	\bar{\zeta} = p_{\LoS}\exp\left(\frac{\rho^2\sigma_{\LoS}^2}{2} - \rho\mu_{\LoS}\right) + p_{\NLoS}\exp\left(\frac{\rho^2\sigma_{\NLoS}^2}{2} - \rho\mu_{\NLoS}\right)~,
\end{equation}
where $\rho = \ln 10/10$. Accordingly, the coverage probability is derived as follows
\begin{align}\label{Eq_SatCov}
	p_\mathrm{c}&(\gamma_\mathrm{o}) = F_{\varphi_\mathrm{o}}(\varphi_\text{m}) - \int_{0}^{\varphi_{\text{m}}}F_\zeta\left(\frac{\gamma_{\mathrm{o}}[\bar{I} + W]}{ P_\mathrm{t}G_\mathrm{t}G_\mathrm{s} l(\varphi)}\right)f_{\varphi_\mathrm{o}}(\varphi)\mathrm{d}\varphi,
\end{align}
where $\gamma_\mathrm{o}$ is the target signal-to-interference and noise ratio, $W$ is the AWGN average noise power. For the detailed derivation, see Appendix~\ref{Appendix_b}. $F_\zeta(\cdot)$ in \eqref{Eq_SatCov} is the CDF of the excess-gain in linear form and is obtained as follows
\begin{align}\label{Eq_Fzeta}
	F_\zeta(x) = \frac{1}{2} +  &\frac{p_{\LoS}(\varphi)}{2}\text{erf}\left[\frac{10\log_{10} x + \mu_{\LoS}}{\sqrt{2}\sigma_{\LoS}}\right]\nonumber\\ &+ \frac{p_{\NLoS}(\varphi)}{2}\text{erf}\left[\frac{10\log_{10} x + \mu_{\NLoS}}{\sqrt{2}\sigma_{\NLoS}}\right].
\end{align}

\section{Uplink Performance Optimization}\label{Sec_Opt}
From \eqref{Eq_SatCov} we can conclude that the performance of the network is predicated on many parameters such as the propagation environment, the number of the satellites, the altitude of the constellation, the effective beamwidth, and the density of ground users. In practical network deployments however, network designers typically have little control on many of these parameters. Nevertheless, designers can effectively tune the beamwidth to maximize the coverage probability even after the deployment of the constellation. In addition, the altitude can be optimized during the design phase of the constellation. Following is a detailed analysis:

\subsubsection{Altitude}
The constellation altitude can significantly affect the performance of the network, whereby increasing the altitude enlarges the satellite footprint and by extension the availability in \eqref{eq_availabilitycdf}. However, a larger satellite footprint lead to elevated levels of aggregate interference due to the increased number of ground users. Moreover, as the altitude increases, the received power reduces and the receiver may not be capable of successfully decoding the received signal. These two opposing factors result in an optimal altitude that maximizes the coverage probability and can be numerically found as follows
\begin{align}\label{Eq_hOptimal}
	h^{\star}(\gamma_{\mathrm{o}},N,\lambda,\psi) =\argmax_{h}\left[p_\mathrm{c}(\gamma_{\mathrm{o}},N,\lambda,h,\psi)\right].
\end{align}
\subsubsection{Effective Beamwidth}
A narrower beamwidth leads to smaller footprint and thus reduces the number of interfering ground users. However, this comes at a cost of reduced satellite availability. Accordingly, an optimal antenna beamwidth that maximizes the coverage probability can be obtained as follows
\begin{equation}\label{Eq_psiptimal}
	\psi^{\star}(\gamma_{\mathrm{o}},N,\lambda,h) =\argmax_{\psi}\left[p_\mathrm{c}(\gamma_{\mathrm{o}},N,\lambda,h,\psi)\right].
\end{equation}
\subsubsection{Joint}
Finally, by jointly optimizing both the altitude and beamwidth, the coverage probability can be significantly enhanced as follows
\begin{equation}\label{Eq_hpsiptimal}
	[h^{\star},\psi^{\star}] =\argmax_{h,\psi}\left[\bar{p}_\mathrm{c}(\gamma_{\mathrm{o}},N,\lambda, h, \psi)\right].
\end{equation}
These three approaches are numerically evaluated based on the integral present in \eqref{Eq_SatCov}. We show that the optimal parameters of the random constellation model perfectly match those obtained from extensive simulations for practical Walker-Star and Walker-delta in Section~\ref{Sec_Res}.

\begin{table}
	\caption{System Parameters}
	\centering
	\begin{tabularx}{3.49in}{l l l}
		\hline\hline \\[-1.5ex]
		Symbol & Value &Definition \\
		\hline\\ [-1.5ex]
		$R_{\tinyEarth}$ & 6371~km & Earth's average radius \\
		$f$ &2~GHz &Center frequency \\
		$P_\mathrm{t}$, $G_\mathrm{t}, G_\mathrm{r}$ & 23, 0, 0~dB & User transmit EIRP, satellite antenna gain \\
		$D, \gamma_{\mathrm{o}}$ & 1\%, -20 dB & User duty cycle, target SINR \\
		$\beta$ &2.3 &LoS probability parameter \\
		$\mu_{\LoS}, \mu_{\NLoS}$ &0, 12~dB\cite{9257490} &LoS/NLoS excess path-loss mean \\
		$\sigma_{\LoS}, \sigma_{\NLoS}$ &2.8, 9~dB\cite{9257490} &LoS/NLoS excess path-loss standard dev. \\
		\hline 
	\end{tabularx}
	\label{Table_Notations}
\end{table}

\section{Simulation and Discussion}\label{Sec_Res}
In this section, we validate the analytical modeling versus the simulation results for the coverage probability and its optimization using both the altitude and beamwidth. To validate the derivation, we consider the Earth to be an ideal sphere with an average radius of 6371~km where the satellites and ground users are both realized using two independent BPPs. Other parameters that correspond to the network are summarized in Table.~\ref{Table_Notations}.

\begin{figure}[!t]
	\begin{subfigure}{\linewidth}
		%\caption{}
		\captionsetup{font=footnotesize,labelfont=footnotesize}
		\centering
		\includegraphics[width=\linewidth]{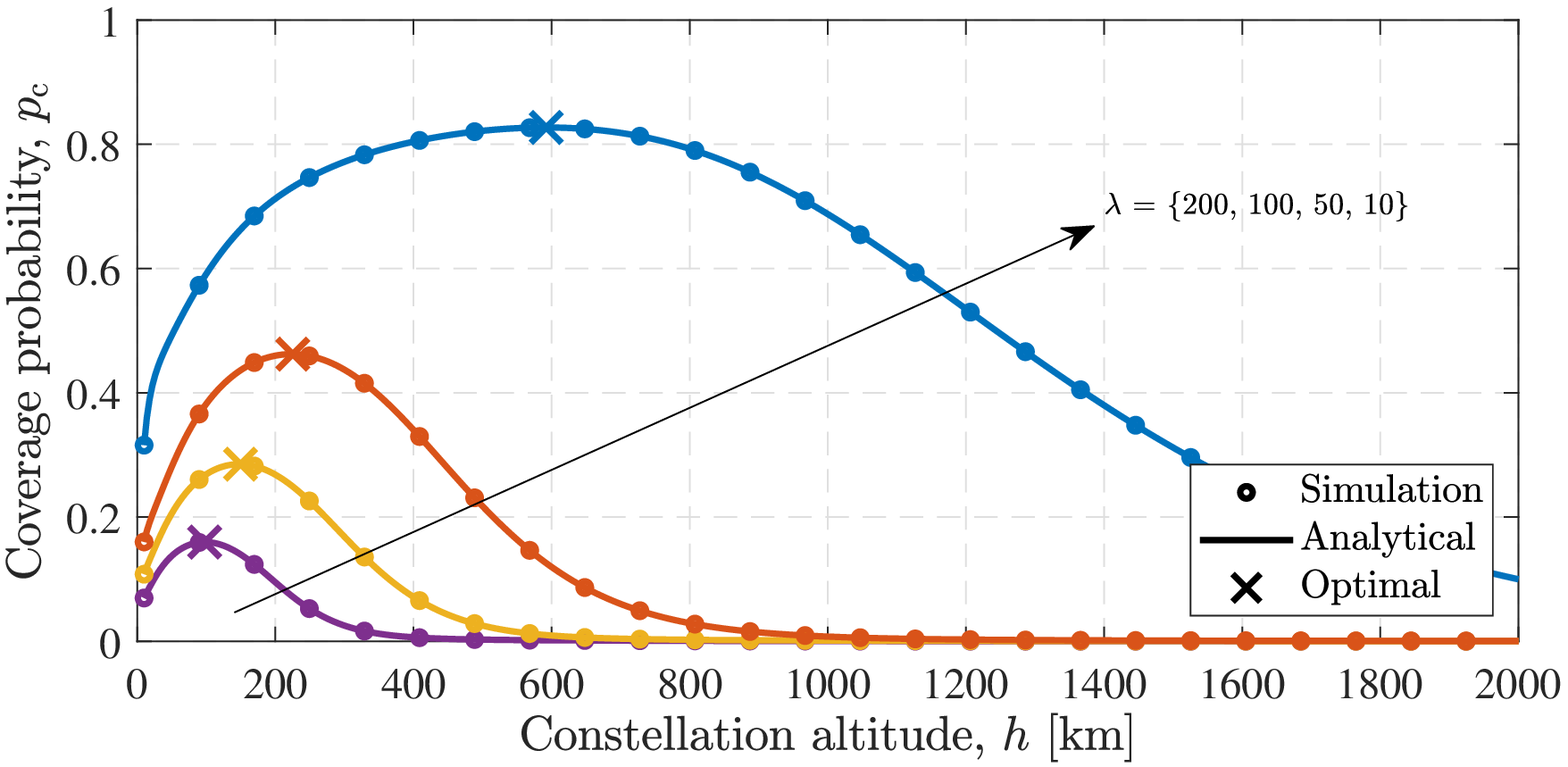}
	\end{subfigure}
	\begin{subfigure}{\linewidth}
		%\caption{}
		\captionsetup{font=footnotesize,labelfont=footnotesize}
		\centering
		\includegraphics[width=\linewidth]{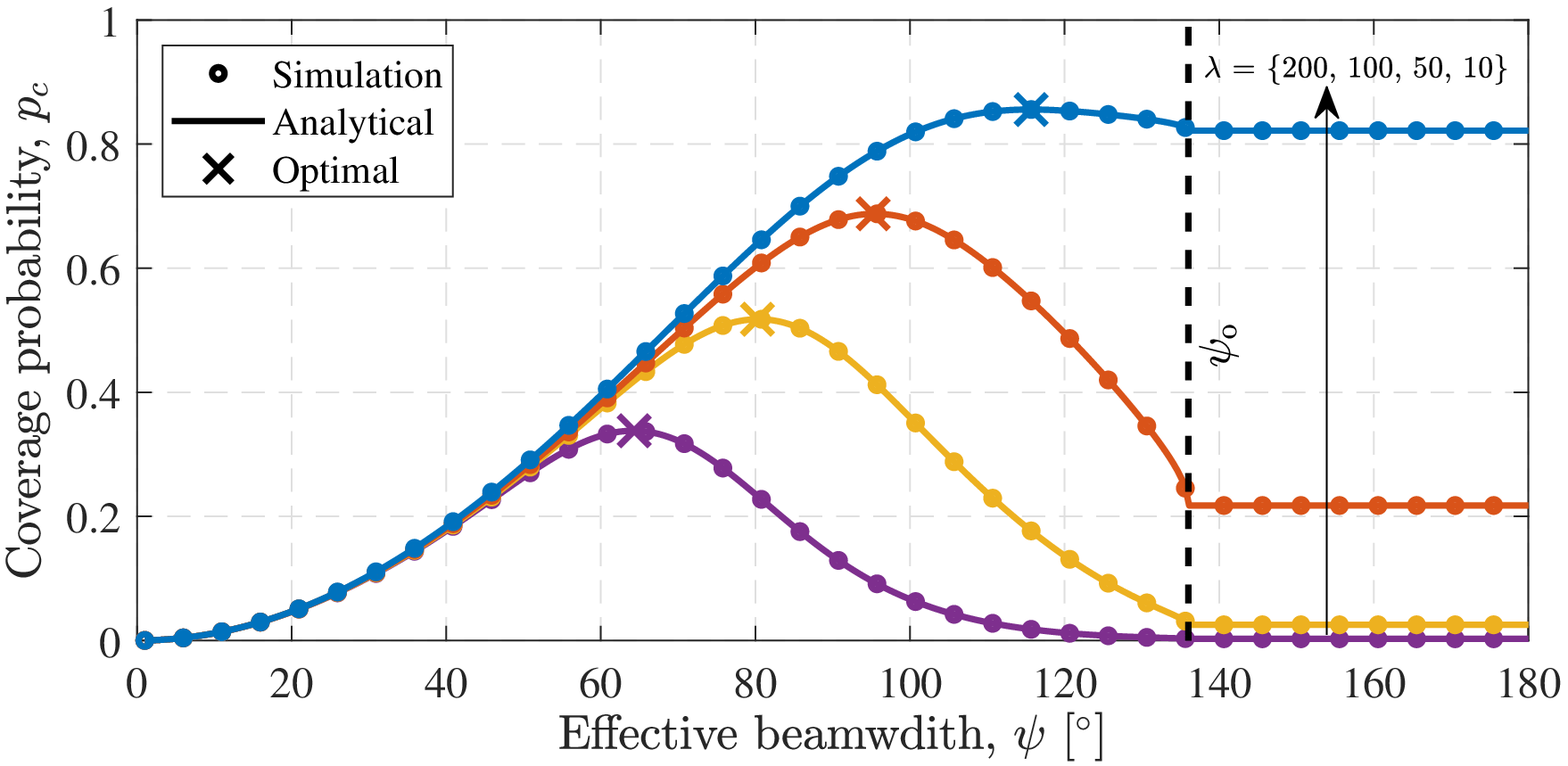}
	\end{subfigure}
	\captionsetup{font=footnotesize,labelfont=footnotesize}
	\caption{Coverage probability for variable (i) altitude at $\psi = $ 90$^\circ$ and (ii) beamwidth at $h = $ 500~km, showcasing the impact of the density of ground users for a constellation size of $N = $ 1000.}
	\label{Fig_CoverageProbability}
\end{figure}
\begin{figure}[!t]
	\captionsetup{font=footnotesize,labelfont=footnotesize}
	\centering
	\includegraphics[width=\linewidth]{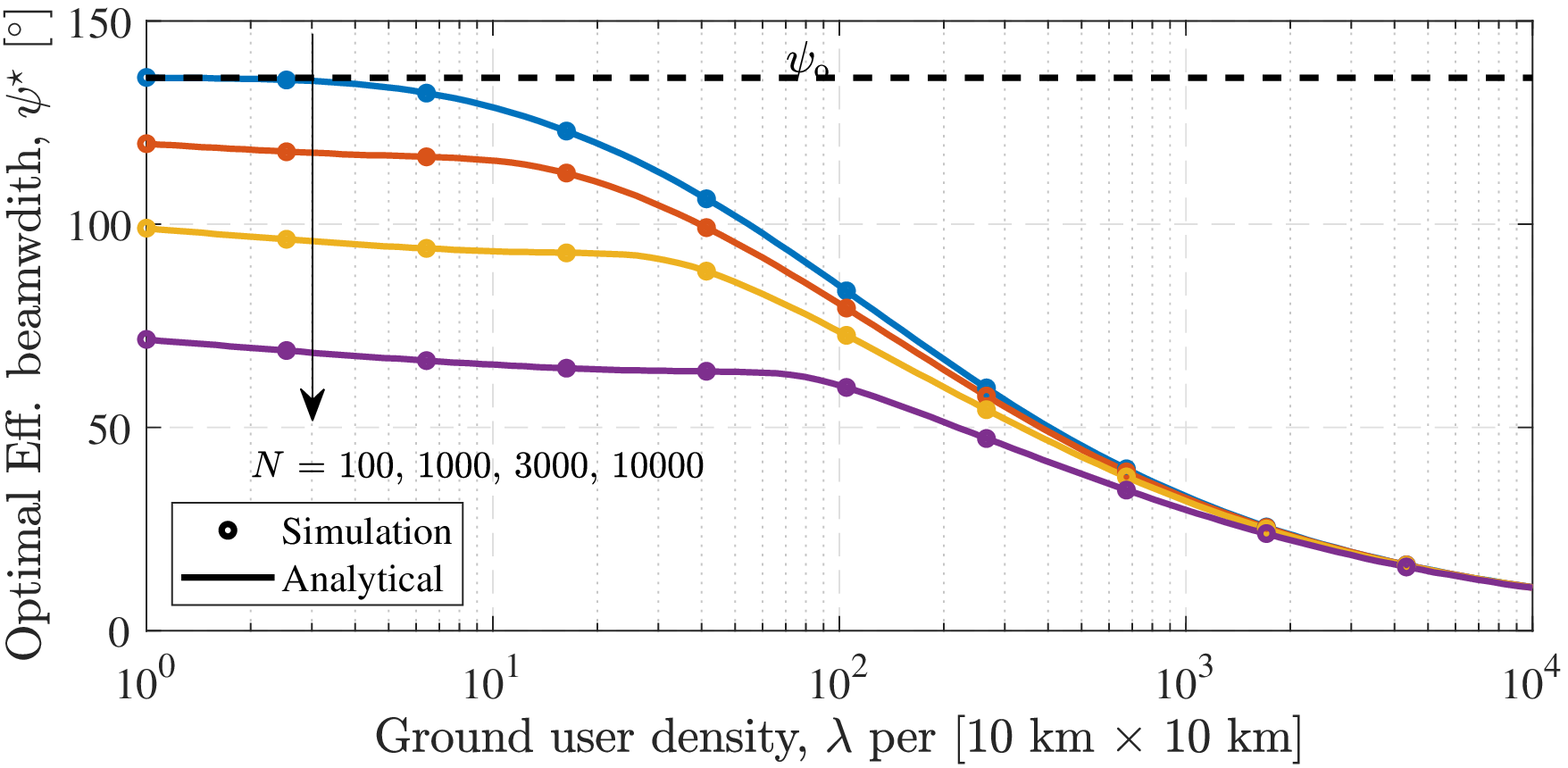}
	\caption{Optimal beamwidth for variable ground user density showcasing the impact of satellite number for $h = $ 500 km.}
	\label{Fig_OptimalBWVarryLambda}
	\footnotesize
\end{figure}

Fig~\ref{Fig_CoverageProbability} provides an insight into the coverage probability as we tune both the constellation altitude (top) and the beamwidth (bottom) separately. The figure shows that the analytical modeling obtained in \eqref{Eq_SatCov} accurately fits the simulation results. The altitude analysis shows that the coverage probability degrades at low altitudes due to the low availability of the satellites. As the altitude increases the coverage improves until it reaches the maximum and then starts degrading rapidly at higher altitudes due to the escalated path-loss. Moreover, as the availability of the satellites increases, so does the footprint area, leading to aggravated levels of interference and degradation in coverage. Similarly, the coverage probability is initially very low for a small effective beamwidth due to the low availability. This however changes as the beamwidth is increased also reaching a maximum before it starts degrading. This is because a large beamwidth leads to increased footprint area and thus exacerbated levels of interference.

Scalability is a key feature for a network that service providers and designers need to account for, whereby more ground users can be easily added to the network. However, for operating satellite constellations, altitude optimization is impractical due to cost and energy limitations making it extremely difficult for satellites to change their altitude after deployment. Nevertheless, optimization is feasible via two methods, (i) launching more satellites or (ii) beamwidth optimization. Adding more satellites is typically difficult mainly due to costs. Hence, beamwidth optimization, highlighted in Fig.~\ref{Fig_OptimalBWVarryLambda}, can accommodate for more ground users by simply capitalizing on adaptive antennas capable of changing their beamwidth on demand, at either the users or satellites end.

\begin{figure}[!t]
	\captionsetup{font=footnotesize,labelfont=footnotesize}
	\centering
	\includegraphics[width=\linewidth]{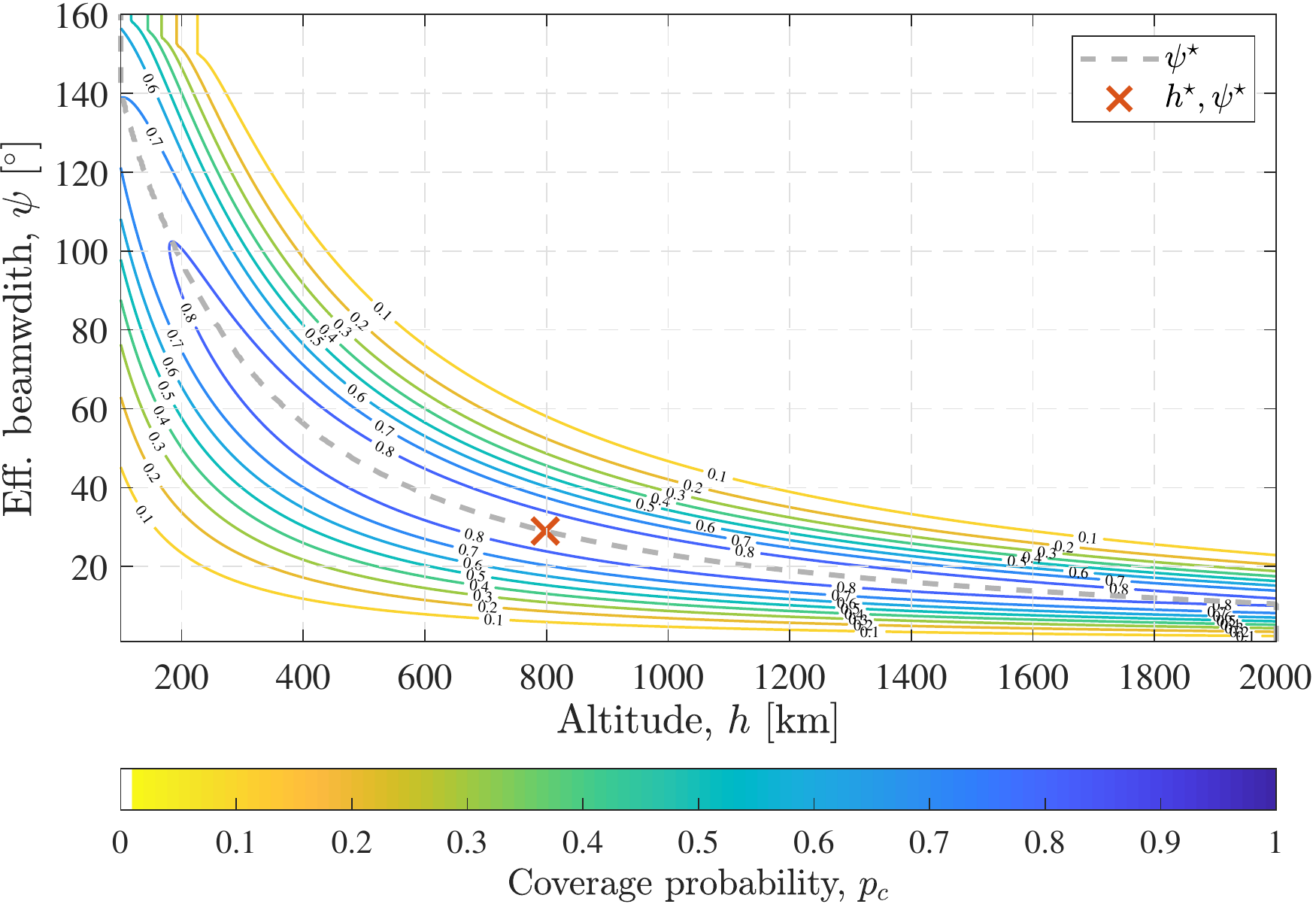}
	\caption{Coverage contour for varying altitude and beamwidth for $N = $ 10,000 and $\lambda = $ 4 per 100 km$^2$. Dashed line is the optimal beamwidth as a function of altitude and the cross is absolute maximum.}
	\label{Fig_Contour}
	\footnotesize
\end{figure}

\begin{figure}[!t]
	\captionsetup{font=footnotesize,labelfont=footnotesize}
	\centering
	\includegraphics[width=\linewidth]{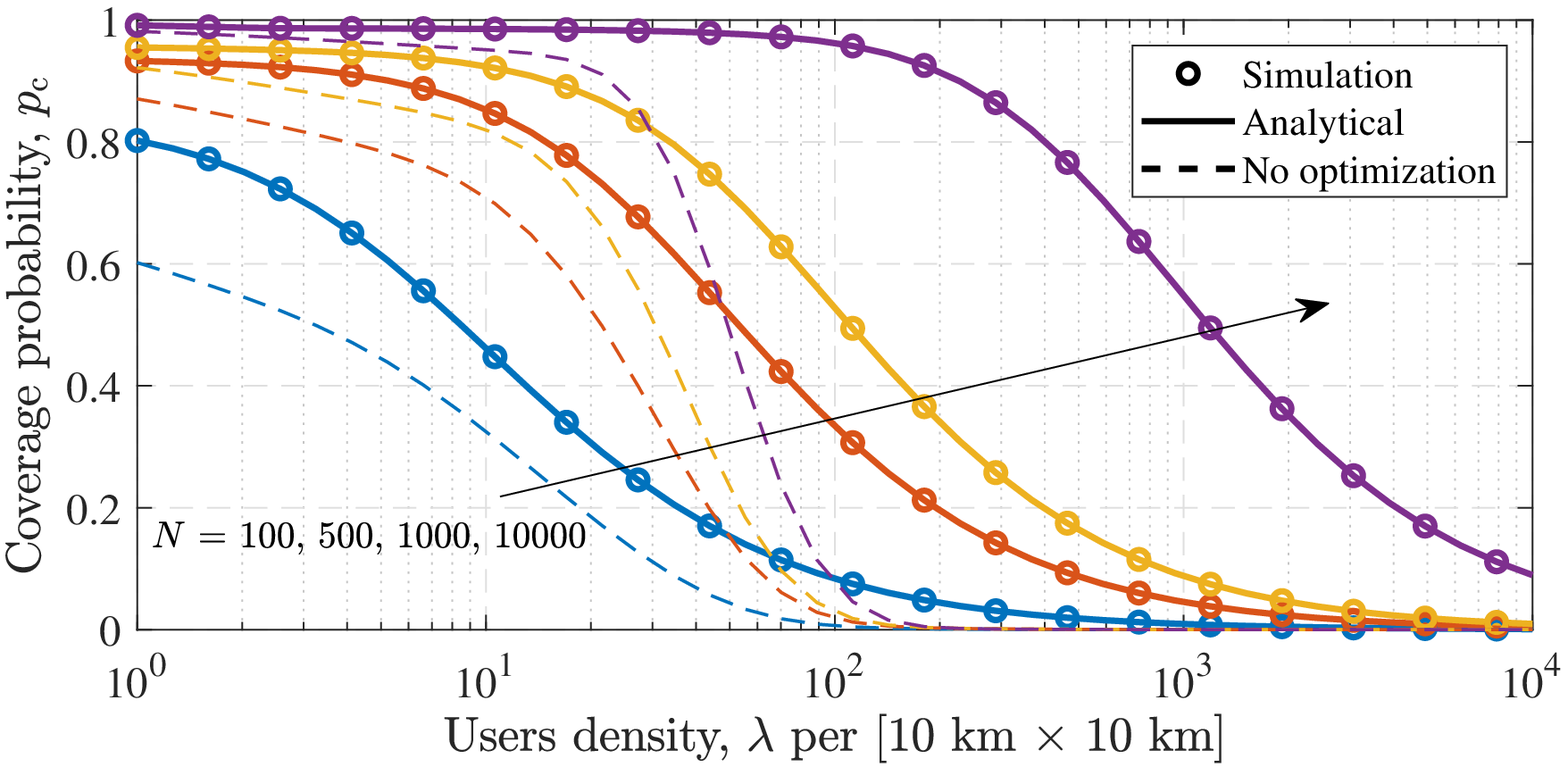}
	\caption{Coverage probability with joint optimization (solid) and without (dashed) for variable ground user density. An altitude of 500 km with an isotropic antenna are assumed for the network without optimization.}
	\label{Fig_All}
	\footnotesize
\end{figure}

On the other hand, for new networks that are planned to be deployed, jointly optimizing both the altitude and beamwidth would produce the best possible coverage probability due to the extra degree of freedom. Fig.~\ref{Fig_Contour} illustrates a contour of the coverage probability as the altitude and beamwidth are jointly varied. The dashed line represents the optimal beamwidth as a function of the altitude whereas the cross represents the most optimal combinations. This optimization enables network designers to develop expansion strategies to accommodate for more devices. Fig.~\ref{Fig_All} underscores the significant coverage probability gain that optimization provides by supporting more ground users without the need to add more infrastructure to the network. Network designers can utilize this framework to unlock more capacity by exploiting link optimization. 

\begin{figure}[!t]
	\captionsetup{font=footnotesize,labelfont=footnotesize}
	\centering
	\includegraphics[width=\linewidth]{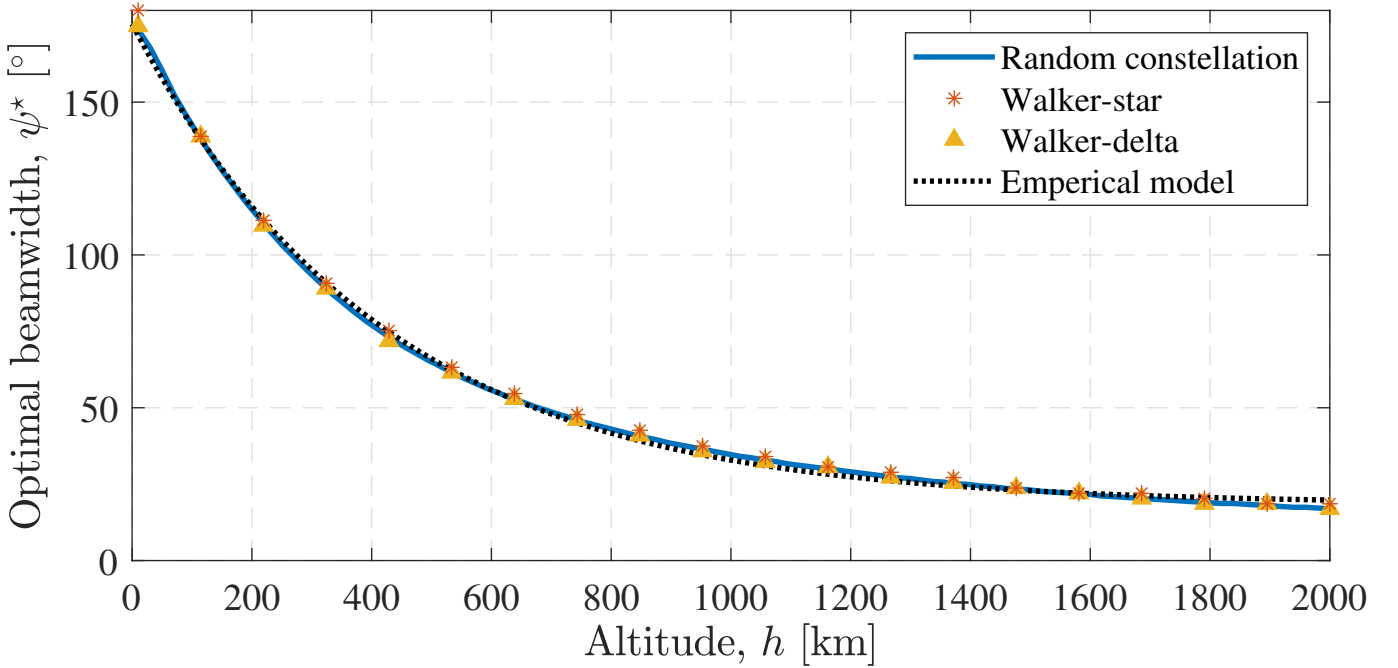}
	\caption{Optimal beamwidth as a function of the altitude for random constellation showing a perfect match with Walker-star and Walker-delta.}
	\label{Fig_Walker}
	\footnotesize
\end{figure}

Furthermore, random constellations typically provide the coverage probability lower bound in comparison to practical constellation deployments such as Walker-star (e.g. SpaceX's Starlink) and Walker-delta (e.g. OneWeb). However, it can provide a close match with practical constellations' optimal altitude and beamwidth as illustrated in Fig.~\ref{Fig_Walker}. Note that the number of planes for both Walker constellations is equal to $\sqrt{N}$ such that the number of satellites per plane is equal to the number of orbital planes. The inclination angle $i$ for Walker-delta and Walker-star constellations are chosen as 86.4$^\circ$ and 53$^\circ$, respectively. Only ground users with latitudes that falls within $i + \varphi_\mathrm{m}$ and $-i - \varphi_\mathrm{m}$ are considered.

\section{Conclusion}\label{Sec_Conc}
Next generation wireless networks are expected to capitalize on dense satellite constellations to provide connectivity for users in rural and remote locations, including massive IoT sensor networks. In this paper, we provide a framework for the optimization of the uplink coverage for satellite constellations by jointly tuning the altitude and the effective beamwidth. The framework utilizes stochastic geometry to develop analytical modeling for the uplink coverage probability. The optimization of random constellations shows a close match to practical satellite networks. This framework enables network designers to rapidly optimize constellation designs and to further develop expansion strategies to cope with increasing user demand.

\appendix
\section{}
\subsection{Average Interference Proof}\label{Appendix_a}
\noindent
We assume that the ground users are within a circular strip area on Earth's spherical surface right underneath the satellite, $f_\mathcal{A}(\varphi)=2\pi R_{\tinyEarth}^2\sin\varphi~\mathrm{d}\varphi$. Accordingly, the average interference is obtained by calculating the expectation over the aggregate interference as follows
\begin{align}
	\bar{I} &= \mathbb{E}_{\Phi}\mathbb{E}_\zeta\left[\sum_{x_i\in\mathcal{A}\backslash x_\mathrm{o}} \kappa P_\mathrm{t}G_\mathrm{t}G_\mathrm{s}l(\varphi_i)\zeta(\varphi_i)\right] \nonumber\\&\stackrel{(a)}{=}  \mathbb{E}_{\Phi}\left[\sum_{x_i\in\mathcal{A}\backslash x_\mathrm{o}} \kappa P_\mathrm{t}G_\mathrm{t}G_\mathrm{s}l(\varphi_i)\bar{\zeta}(\varphi_i)\right] \nonumber\\
	&\stackrel{(b)}{=} \lambda \kappa P_\mathrm{t}G_\mathrm{t}G_\mathrm{s} \int_{0}^{\varphi_\mathrm{m}}  l(\varphi)\bar{\zeta}(\varphi)f_\mathcal{A}(\varphi)~\mathrm{d}\varphi~,
\end{align}
where $(a)$ stems from the fact that $\zeta$ is independent from the users and thus its expectation can be taken inside the summation and $(b)$ stems from invoking Campbell's theorem of sums for a stationary point process given as follows~\cite{HaenggiBook}
\begin{align}
	\mathbb{E}_{\Phi}\left[\sum_{x\in\Phi}f(x)\right] = \lambda\int_{\mathbb{R}^d}f(x)~\mathrm{d}x.
\end{align}

\subsection{Coverage Probability Proof}\label{Appendix_b}
\noindent
The signal-to-interference and noise ratio (SINR) of a ground user is defined as
\begin{equation}
	\gamma = \frac{P_\mathrm{r}}{I + W} = \frac{ P_\mathrm{t}G_\mathrm{t}G_{\mathrm{s}}l(\varphi)\zeta(\varphi)}{I + W}.
\end{equation}
Accordingly, the coverage probability is the probability that the SINR is larger than a certain threshold as follows
\begin{align}\label{Eq_SatCovApp}
	p_\mathrm{c}(\gamma_\mathrm{o}) &= \mathbb{P}\left(\gamma > \gamma_{\mathrm{o}}\right) 
	= \mathbb{P}\left(\frac{P_\mathrm{r}}{I + W} > \gamma_{\mathrm{o}}\right) \nonumber\\
	&= \mathbb{E}_{\varphi_\mathrm{o}}\left[\mathbb{P}\left(\zeta > \frac{\gamma_{\mathrm{o}}[\bar{I} + W]}{ P_\mathrm{t}G_\mathrm{t}G_\mathrm{s} l(\varphi_\mathrm{o})}\right)\right] \nonumber\\
	&= \mathbb{E}_{\varphi_\mathrm{o}}\left[1 - F_{\zeta}\left( \frac{\gamma_{\mathrm{o}}[\bar{I} + W]}{P_\mathrm{t}G_\mathrm{t}G_\mathrm{s} l(\varphi_\mathrm{o})}\right)\right] \nonumber\\
	&=  \int_{0}^{\varphi_{\text{m}}}\left[1 - F_\zeta\left(\frac{\gamma_{\mathrm{o}}[\bar{I} + W]}{P_\mathrm{t}G_\mathrm{t}G_\mathrm{s} l(\varphi)}\right)\right]f_{\varphi_\mathrm{o}}(\varphi)~\mathrm{d}\varphi.
\end{align}

\ifCLASSOPTIONcaptionsoff
\newpage
\fi
\bibliographystyle{IEEEtran}
\bibliography{SatelliteOptimizationUplink}

\end{document}